\documentclass{article}
\usepackage{PRIMEarxiv}
\usepackage[utf8]{inputenc} 
\usepackage[T1]{fontenc}    
\usepackage{palatino,eulervm}
\usepackage{hyperref}       
\usepackage{url}            
\usepackage{booktabs}       
\usepackage{amsfonts}       
\usepackage{nicefrac}       
\usepackage{microtype}      
\usepackage{lipsum}
\usepackage{fancyhdr}       
\usepackage{graphicx}       
\graphicspath{{media/}}     

\title{SVD Factorization for Tall-and-Fat Matrices on Parallel Architectures}

\author{
  Burak Bayramlı \\
  İstanbul, Turkey\\
  \texttt{burakbayramli.github.io} 
}

\begin{document}

\maketitle

\begin{abstract}
We demonstrate an implementation for an approximate rank-k SVD factorization,
combining well-known randomized projection techniques with previously known
paralel solutions in order to compute steps of the random projection based SVD
procedure. We structure the problem in a way that it reduces to fast computation
around $k \times k$ matrices computed on a single machine, greatly easing the
computability of the problem. The paper is also a tutorial on paralel linear
algebra methods using a plain architecture without burdensome frameworks.
\end{abstract}

\keywords{Parallel \and Concurrency \and Big Data}

\section{Introduction}

Parallelization for Big Data type problems where we are presented with many
rows, reaching upwards of billions, large (but not as many) columns, can be
achieved by processing the input file line by line where the process
contributes, furthers the computation at each line, accumulating a small result
in memory, or written out to disk as a line, at each iteration. Parallel
implementation can be possible by assuming each process on each machine has
access to a large file, can process different parts of this big file, either
through copies of that file being in each machine, or through a shared file
server on a fast Ethernet network. As most number crunching applications are CPU
bound, potential IO blockages are deemed as less of an issue for such problems.

Many methods in Linear Algebra can be reframed this way where line-by-line
computation is pursued, the processing starts with first line, computes, and
goes further down, furthering the computation each step after which results from
each process are added up.

\section{Linear Algebra Basics}

\subsubsection{Singular Value Decomposition}

Computing SVD on an $m \times n$ matrix where $m$ is large can be
taxing. However by making use of $A$'s representation as $A = U \Sigma V^T$, and
computing $A^TA$, we realize \cite{zadeh},

$$
A^TA = V \Sigma^2 V^T 
$$

The righthand side has reduced to an eigenvector calculation. Therefore by
computing $A^TA$ we are computing $V \Sigma^2 V^T $ on which we can run
eigenvector calculation to obtain the $V$ vector. The approach can be more
optimal because if $A$ is $m \times n$ where $n << m$, therefore $A^TA$
dimensions has to be $n \times n$ a much smaller matrix that can fit in memory.
This way SVD calculation of a large $A$ is reduced eigenvector calculation of a
smaller $A^TA$.

Calculating $U$ is done as follows,

$$
A = U \Sigma V^T \to U = A V \Sigma^{-1}
$$

\subsubsection{Incremental $A^T A$}

We have reduced incremental, parallel computation of SVD to parallel computation
of $A^T A$ which, in turn, can be pursued by noticing $A^T A$ is easily computed
through the outer product of $i$'th row of the matrix, $A_{i}$ with itself, and
summing the results.

$$
A^T A = \sum_{i} A_i \times A_i
$$

Summation is commutative, therefore outer products from each parallel process,
per row, can be added together, no matter the order, first at each node as a
local sum, than as a global sum once the processing of each node is finished.

Simple Python code demonstration,

\begin{verbatim}
A = [[1,2,3],
     [3,4,5],
     [4,5,6],
     [6,7,8]]
A = np.array(A)

s = np.zeros((3,3))
for i in range(4):
  s = s + np.outer(A[i,:],A[i,:])
print (s)    
\end{verbatim}

\begin{verbatim}
[[ 62.  76.  90.]
 [ 76.  94. 112.]
 [ 90. 112. 134.]]
\end{verbatim}

\subsubsection{Multiplication}

Let's assume we have matrices $A$, and $B$. Naive matrix multiplication has
$O(n^3)$ complexity for $n \times n$ matrices sizes for both $A$ and $B$. First
stab at the incremental multiplication in pseudocode is as follows,

\begin{verbatim}
for(int m = 0; m < M; m++) {
    for(int k = 0; k < K; k++) {
        for(int n = 0; n < N; n++) {
            C[m][n] += A[m][k]*B[k][n];
        }
    }
}
\end{verbatim}

\begin{figure}[h]
  \centering
  \includegraphics[width=20em]{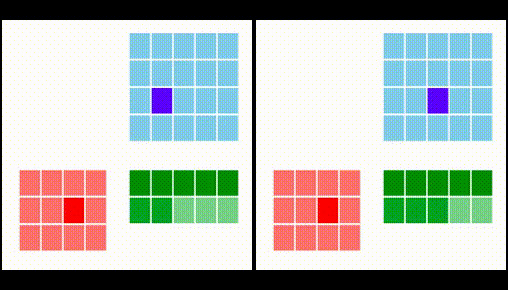}
  \caption{Two Steps of a Row Based Multiplication Process}
  \label{fig:mult1}
\end{figure}

In the case we are iterating $A$ and $B$ incrementally it needs to be noted for
large $B$ matrices, $B$ has to be iterated row by row, top to bottom for each
row of $A$. That is why complexity is $O(n^3)$.

For cases where $B$ can fit in memory however, we can simplify the algorithm.
In this case, each row $c_i$ of multiplication result $C$ is nothing but a row
of A times the entirety of $B$ which gives us $a_i \cdot B$. Simple Python code
can demonstrate such a multiplication where rows of $A$ are multiplied with all
of $B$.

\begin{verbatim}
A = np.array([[1,2,3],[2,2,2]])
B = np.array([[3,4,5],[1,1,1],[2,2,2]])
print ('All C')
print (A.dot(B))
print ('C row 1',A[0,:].dot(B))
print ('C row 2',A[1,:].dot(B))
\end{verbatim}

\begin{verbatim}
All C
[[11 12 13]
 [12 14 16]]
C row 1 [11 12 13]
C row 2 [12 14 16]
\end{verbatim}

In this paper we assumed matrix $B$, in dimensions $n \times k$, can be brought
into memory completely. The operation as seen above will be computed for each
row of $A$ and the results will be summed.

\subsection{Random $B$ Multiplication}

For large scale SVD reducing the size of $A$ through random projection obtaining
an approximation is possible. This approximation works for the reason that
projecting points to a random subspace preserves distances between points, or in
detail, projecting the n-point subset onto a random subspace of $O(\log
n/\epsilon^2)$ dimensions only changes the interpoint distances by $(1 \pm
\epsilon)$ with positive probability \cite{gupta}. It is also said that
projected matrix $Y$ is a good representation of the span of $A$.

The multiplication shown previously was a general-purpose approach, which can be
used to multiply $A$ with any $B$ where $B$ is a random matrix. For the case of
random projection we can make certain improvements to the logic. For random
projection the only requirement for $B$ is its rows are generated from normal
$N(0,1)$ distribution. Since we know pseudonumber generation is deterministic
running in constant time, we could regenerate rows of $B$ as needed without
keeping a large $B$ in memory, or constantly rescan it from top to bottom for
each row of $A$. We simply need to make sure the {\em same} random $B$ rows are
used each time, we can easily guarantee that by setting the same seed value on
the generator for each new row of $A$. In simple Python code,

\begin{verbatim}
A = np.array(range(0,20)).reshape(5,4)
Y = []; k = 2 # reduction dimension
for i in range(A.shape[0]):
    np.random.seed(0)
    s = np.zeros(k)
    for e in A[i, :]: s += e*np.random.normal(0,1,k)
    Y.append(s)
Y = np.array(Y)
\end{verbatim}

\section{Implementation}

Big Data became possible largely thanks to Map-Reduce architectures which
represent splitting (mapping) and grouping (reducing) concepts logically and
expose them as the only interface for a programmer to worry about, while in the
background directing data pieces produced by mapping and reducing to appropiate
seperate nodes to achieve concurrency. A sample process is seen in
Figure~\ref{fig:mapreduce1},

\begin{figure}[h]
  \centering
  \includegraphics[width=20em]{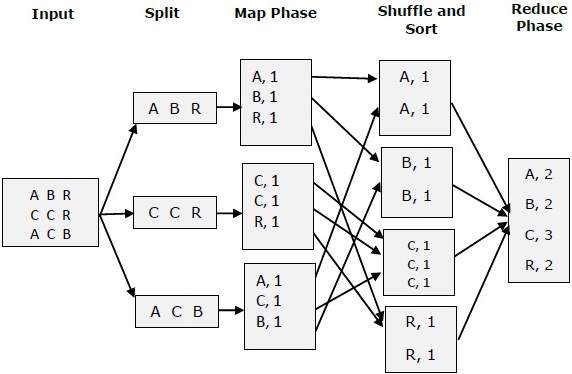}
  \caption{Example of Map-Reduce}
  \label{fig:mapreduce1}
\end{figure}

Our proposal is a simpler, so-called Split-Process architecture. Each process
has access to the large input file, is able to skip ahead to any row of that
file, distribution is done on the basis of processing a pre-decided subsets of
that data.

\begin{figure}[h]
  \centering
  \includegraphics[width=15em]{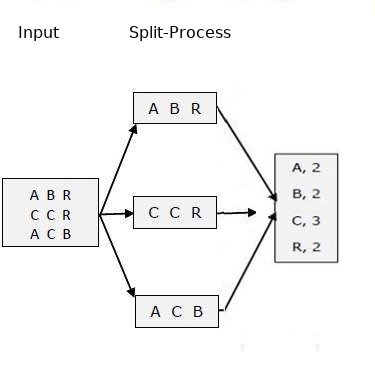}
  \caption{Split-Process}
  \label{fig:mapreduce1}
\end{figure}

The nature of the computation described previously fits perfectly with this
approach. With four processes and 1000 lines to process, each process
$i=1,2,3,4$ can be directed to focus on their portion of the file, first can
take rows $1,..,250$, the next can take $251,..,500$, so on.

Here we present the main processor of the code, called \verb!split_process!
which can determine line beginning and end points in terms of seek byte
locations of a given file, and by inspecting the \verb!workobj.ci! will
jump ahead toward the chunk and start reading the file line by line, and
feed them into \verb!workobj.exec!. In an object-oriented design will allow each
\verb!workobj! to know how to handle an input line, will either accumulate
results from it or write an output itself to another file.

\begin{verbatim}
import os, numpy as np

def split_process(file_name,N,workobj):
    file_size = os.path.getsize(file_name)
    beg = 0
    chunks = []
    for i in range(N):
        with open(file_name, 'r') as f:
            s = int((file_size / N)*(i+1))
            f.seek(s)
            f.readline()
            end_chunk = f.tell()-1
            chunks.append([beg,end_chunk])
            f.close()
        beg = end_chunk+1
    c = chunks[workobj.ci]
    with open(file_name, 'r') as f:
        f.seek(c[0])
        while True:
            line = f.readline()
            workobj.exec(line)
            if f.tell() > c[1]: break
        f.close()
        workobj.post()
\end{verbatim}

The results, if they can be accumulated in memory, can be kept on
\verb!workobj!. Once a chunk is finished (all its rows are visited) then
\verb!split_process! will call \verb!workobj.post! which can handle disk output,
cleaning up operations.

The job classes for $A^T A$, $AB$ and random projection are given below.

\subsection{Calculation of $A^T A$}

\begin{verbatim}
class ATAJob:
    def __init__(self,D,ci):
        self.C = np.zeros((D,D))
        self.ci = ci
    def exec(self,line):
        tok = line.split(';')
        vec = np.array([float(x) for x in tok])
        self.C = self.C + np.outer(vec, vec)
    def post(self):
        outfile = "/tmp/C-%d.csv" % self.ci
        np.savetxt(outfile, self.C, delimiter=';',fmt='%1.6f')
\end{verbatim}

\subsection{Multiplication of $A$ and $B$}

\begin{verbatim}
class MultJob:
    def __init__(self,ci,bfile):
        self.afile = ""
        self.B = np.loadtxt(bfile,delimiter=';')
        self.ci = ci
        cname = "%s/C-%d.csv" % (os.path.dirname(afile), self.ci)
        self.outfile = open(cname, "w")        
    def exec(self,line):        
        vec = np.array([np.float(x) for x in line.strip().split(";")])
        vec = np.reshape(vec, (len(vec),1))
        res = (vec * self.B).sum(axis=0).tolist()  
        res = ";".join(map(str, res))
        self.outfile.write(res)
        self.outfile.write("\n")
        self.outfile.flush()
    def post(self):
        self.outfile.close()
\end{verbatim}

\subsection{Random Projection}

\begin{verbatim}
class RandomProjJob:
    def __init__(self,ci):
        self.ci = ci
        self.k = 7
        self.outfile = open("/tmp/Y-%d.csv" % self.ci, "w")        
    def exec(self,line):
        s = np.zeros(self.k)
        toks = line.strip().split(';')
        row = np.array([np.float(x) for x in toks[1:]])
        # degisik veri parcalari degisik rasgele matrisler uretsin
        np.random.seed(0) 
        for elem in row: s += elem*np.random.normal(0,1,self.k) 
        s = ";".join(map(str, s))
        self.outfile.write(s)
        self.outfile.write("\n")
        self.outfile.flush()
        
    def post(self):
        self.outfile.close()
\end{verbatim}

\section{Conclusion}

We demonstrated easy-to-scale parallelization approach that can be used on
different types of linear algebra problems. A final note here is that the random
projection technique, though presented as helping SVD, can also be used in place
of SVD \cite{lu} as preserving distances between projected rows is useful for
any similarity calculation method typically using cosine or Euclidian methods.

\bibliographystyle{unsrt}

\begin{thebibliography}{1}

\bibitem{gleich}
Gleich, Benson, Demmel, \emph{Direct QR factorizations for tall-and-skinny
  matrices in MapReduce architectures}, {\tt arXiv:1301.1071 [cs.DC]}, 2013

\bibitem{halko}
N.~Halko, \emph{Randomized methods for computing low-rank approximations of
  matrices}, University of Colorado, Boulder, 2010

\bibitem{gupta}
S.~Dangupta, A.~Gupta \emph{An Elementary Proof of a Theorem of Johnson and
  Lindenstrauss}, Wiley Periodicals, 2002

\bibitem{kurucz}
M.~Kurucz, A. A.~Benczúr, K.~Csalogány, \emph{Methods for large scale SVD with
missing values}, ACM, 2007

\bibitem{zadeh}
Zadeh, \emph{CME 323: Distributed Algorithms and Optimization, Lecture 17}, 
\url{https://stanford.edu/~rezab/classes/cme323/S17/}

\bibitem{agrawal}
Agrawal, \emph{Matrix Multiplication: Inner Product, Outer Product and Systolic Array},
\url{https://www.adityaagrawal.net/blog/architecture/matrix_multiplication}

\bibitem{lu}
Lu, \emph{On Low Dimensional Random Projections and Similarity Search},
    \url{https://www.researchgate.net/publication/221615011_On_Low_Dimensional_Random_Projections_and_Similarity_Search}

\end{thebibliography}

\end{document}